# Bipolar resistive switching characteristics of poly(3,4-ethylene-dioxythiophene): poly(styrenesulfonate) thin film

Hu Young Jeong [a, b], Jong Yun Kim [a, c], Tae Hyun Yoon [c], Sung-Yool Choi [a,*]

[a] *Convergence Components & Materials Research Lab, ETRI, Daejeon 305-700, Korea*
[b] *Department of Materials Science and Engineering, KAIST, Daejeon 305-701, Korea*
[c] *Department of Chemistry, Hanyang University, Seoul 133-701, Korea*

**Abstract**

We investigated the reversible resistive switching of poly(3,4-ethylenedioxythiophene):polystyrenesulfonate (PEDOT:PSS) thin films sandwiched between Al electrodes. The J-V sweep curve showed a hysteretic behavior which depends on the polarity of the applied voltage bias. From the analysis of I-V curves, it was revealed that the charge transport through the junction was governed by the bulk space-charge-limited conduction (SCLC) model. Using transmission electron microscopy (TEM) analysis, it was confirmed that the initial high resistance state of PEDOT:PSS films is related with the segregation of $PSS^-$ chains induced by redox reaction between a Al metal electrode and PEDOT:PSS film. Positive space charges present on the top region of PEDOT:PSS films can be proposed as a possible trap centers of electron trapping and detrapping process.

---

[*] Electronic Mail: sychoi@etri.re.kr

## 1. Introduction

The resistance switching phenomena of organic thin films sandwiched between two metal electrodes have been intensively studied in several systems of active layer [1], i.e. organic/metal nanocluster/organic triple-layer [2], polymer/metal nanoparticles composite layer [3], and all-organic donor-acceptor system [4] for the potential organic nonvolatile memory applications. Among these organic materials, poly(3,4-ethylene-dioxythiophene): polystyrenesulfonate (PEDOT:PSS) has recently attracted considerable attention due to versatile electrical switching characteristics such as write-once-read-many times (WORM) memory switching in ITO/PEDOT:PSS/Au devices [5], bipolar resistive switching (BRS) in ITO/PEDOT:PSS/Al devices [6], and unipolar resistive switching (URS) in Au/PEDOT:PSS/Au devices [7]. Although it is generally explained that the electronic conductivity of PEDOT:PSS could be modified by the formation and destruction of current paths by the reduction and oxidation of PEDOT chains in PEDOT:PSS thin films [5-7], few works have focused on the interface reaction between active metal electrode and bulk organic thin films.

In this study, we investigated the reversible resistive switching of PEDOT: PSS thin films sandwiched between Al metal electrodes. The different current-voltage (I-V) characteristic with Al/PEDOT:PSS/Al device reported by Kim et al. [8] was observed, and the bipolar switching behavior of PEDOT:PSS thin polymer films is described through the space charges induced by interface reaction between top Al metal electrode and bulk PEDOT:PSS thin films.

## 2. Experimental details

To fabricate Al/PEDOT:PSS/Al devices, a 70nm thick Al bottom electrode was deposited on Si/SiO$_2$ (3000 Å) substrate by thermal evaporation method. The PEDOT: PSS aqueous solution (PH 500, H.C. Starck Inc.; chemical structure as shown in the left inset of Fig. 1) was spun on the Al bottom electrode at 500 rpm for 10 s, then at 2000 rpm for 60 s. After spin coating, the film was dried on a hot plate at 120 ℃ for 40 min in ambient air. The thickness of the films was determined to be about 70nm using a spectroscopic ellipsometer and transmission electron microscope (TEM). The sheet resistance of 2.5 MΩ/□ was measured by 4-probe measurement. Al top electrode of 70 nm thickness was deposited on the top surface of the organic film through a metal shadow mask with square patterns of 200 μm × 200 μm areas. The right inset of Fig. 1 shows the schematic diagram of Al/PEDOT:PSS/Al device. The current-voltage characteristics were measured using Keithley 4200 semiconductor characterization system in ambient air at room temperature. The structural and chemical characterizations were carried out with transmission electron microscopy (TEM) on the cross-sectional specimen prepared by a conventional method. The cross-sectional TEM images of Al/PEDOT:PSS/Al samples were obtained using 300 kV JEOL JEM 3010 with 0.17 nm point resolution. Annular dark filed (ADF) scanning TEM (STEM) image and point energy dispersive x-ray spectroscopy (EDS) data were obtained using field emission transmission electron microscope (FETEM) (Tecnai G2 F30) equipped with STEM mode and an EDS analyzer.

## 3. Results and discussion

Fig. 1 shows the typical current density-voltage (J-V) characteristics of a single layer PEDOT:PSS thin film device illustrated as a semilogarithmic plot. The curve was swept by applying a bias voltage to the top electrode (Al) and grounding the bottom electrode (Al). When the voltage applied to pristine device with negative voltage direction, the device initially had high resistance state (HRS). PEDOT:PSS films are well known as one of the most common electrically conducting organic polymers [9]. However in our system, the PEDOT:PSS films exhibited an insulating property, having a low electrical conductivity of 0.14 S/cm measured in the ohmic region of Fig. 1. This might be due to no addition of dielectric solvent such as dimethyl sulfoxide (DMSO) [10] and interface layer present on both Al metal electrodes. Interestingly, the J-V curve also shows pronounced hysteretic and asymmetric behavior. When the negative voltage increased up to -5V, the gradual increase of current density occurred at the voltage of around -3V. The voltage was then scanned from -5V to 0V, the different low resistance state (LRS) emerged. The LRS was retained until the applied voltage reached ~3V. The reversible hysteretic current-voltage characteristic was repeatedly observed during several voltage sweeps.

To examine the possible switching mechanism of PEDOT: PSS thin films, we investigated the I-V curve on a double log scale. Fig. 2 shows the double logarithmic plots of the I-V curve for the negative and positive voltage sweep regions, respectively. In the figure, there are four distinct regimes marked with dashed circles. In the first negative increasing scan, shown in Fig. 2a, current is linearly proportional to the voltage, corresponding to the ohmic conduction ($I \propto V$) dominated by thermally stimulated free electrons, at low voltage region (<0.1V). Then the current follows a square dependence on voltage: $I \propto V^2$, which corresponds to Child's law region [11]. In this region, the density of injected charge carriers is much higher than that of thermally generated free carriers inside PEDOT:PSS films. When the applied voltage reaches to a certain threshold voltage $V_{th}$( ~ 3 V), the current increases rapidly. According to SCLC theory [11], a less abrupt increase of current is related with the filling process of traps in the polymer layer. In higher-voltage region $(V > V_{th})$, $I \propto V^2$ is observed again. In the negative voltage-decreasing sweep, the current decreased as $I \propto V^2$ ,described as trap-free SCLC regime, followed by $I \propto V$ preserving the low resistance state. On the other hand, in the positive bias region, as shown in Fig. 2b, a similar SCLC behavior of low resistance state is shown. However, in the high-voltage region of increasing sweep, the current decreases significantly between 3V and 4V and then increases again. This is called as negative differential resistance phenomenon (NDR). It is verified that the device is restored to the high resistance state (HRS) from the decreasing positive scan. Thus, the I-V characteristics of this system can be clearly explained by the bulk trap-controlled space-charge-limited-current (SCLC) mechanism, as previously reported in various organic and inorganic materials [12,13].

For the clear characterization of interface layer between Al metal electrodes and PEDOT:PSS polymer thin film, the TEM analyses such as high-resolution TEM (HRTEM) and STEM point EDS analyses were performed. Fig. 3a shows a cross-sectional bright-field (BF) TEM image of the Al/PEDOT:PSS/Al structure with three individual layers (the Al bottom electrode, the PEDOT:PSS organic layer, and the Al top electrode). It is shown that the highly uniform PEDOT:PSS polymer layer with amorphous phase, having no lattice images, was deposited on the Al bottom electrode with a good roughness due to slow evaporation rate (0.8 Å/sec). Note that the amorphous interface layers with a dark contrast are observed at the vicinity of both Al metal electrodes. It is easily accepted that very thin Al native oxide layer with a 1~2 nm thickness already existed before spin-coating process due to strong oxidization of Al metal. On the other hand, the top interface layer (~4nm), thicker than bottom interface layer, was formed after the aluminum thermal evaporation process. It is assumed that this layer was newly created by the reaction between Al metal electrode and PEDOT:PSS thin film. It was already reported by many groups that bistability in pure organic thin films is due to the presence of a native oxide layer at the interface between the organic layers and Al electrode during Al deposition, focusing on the pure native oxide thin layer not the interaction [14,15]. However, it is suggested that top interface layer of our Al/PEDOT:PSS/Al system composed of not only Al and O composition, but also other organic materials, on the basis of the broader thickness from above HRTEM images. It is not likely to form pure Al native oxide with around 4~5nm thickness.

To confirm above postulation, we performed a STEM point EDS analysis. Fig. 4a shows annular dark field STEM image of Al/PEDOT:PSS/Al stack on the $SiO_2$ substrate showing atomic Z-contrast. The Al metal with high atomic number shows the brighter contrast. Fig.3b-d shows point EDS data obtained from the three points marked with red circles. These point-analyzed spectra were collected in a 20s live time using a 0.7nm probe size. Because the electron beam broadening is inevitable in the point STEM EDS measurement, it is more reasonable to relatively compare data. The spectra represent the chemical composition of three different regions in the PEDOT:PSS thin film; top interface region, bulk region and bottom interface region. There are three notable findings. First, top interface layer, shown in the HRTEM image, contains organic components (C, O, and S) as well as Al, meaning that the chemical reaction between Al and PEDOT:PSS occurred. Second, the penetration of Al into the active PEDOT:PSS layer is also possible during evaporation process, as confirmed in visible EDS spectra of Al found in the region of PEDOT:PSS active layer. Finally, the relative amounts of C, O and S compositions in the top interface region are larger than bulk and bottom interface region. It is proposed that the compositional inhomogeneity of the active PEDOT:PSS thin film occurred during Al metal deposition.

Xu et al. [16] reported that $PSS^-$ rich PEDOT:PSS thin film showed the irreversible conductivity decrease in the presence of a large current. It is also suggested that the excess $PSS^-$ chains migrate locally toward the electrode under the electric field, resulting in the segregation of the conducting PEDOT:PSS particles and the insulating PSS⁻ chains [16]. Based on this report and above TEM results, the bipolar resistive switching mechanism of our Al/PEDOT:PSS/Al devices can be described schematically in Fig. 5. The PEDOT:PSS (1:2.5; weight ratio) thin film deposited on Al bottom electrode has a granular morphology [17], that is, conducting PEDOT:PSS particles are embedded in a PSS⁻ - rich environment, as shown in Fig. 5a. When the top Al electrode deposition started in the thermal evaporation chamber and aluminum atoms attached to the surface of the PEDOT:PSS film, they react with organic molecules containing carbon, oxygen and sulfur, making itself oxidized. The oxidation state ($Al^{3+}$) strongly attracts the insulating PSS- chains mainly distributed in the active organic film, resulting in initial HRS of our device. The bulk positive space charges induced by the segregation of PSS⁻ chains plays a major role in electron charge trapping /detrapping process of SCLC conduction system.

## 4. Conclusions

In conclusion, we described the reversible resistive switching characteristics of PEDOT:PSS organic thin films deposited by spin-coating. The pristine Al/PEDOT:PSS/Al devices had a low conductivity due to no addition of DMSO solvent and the interface layers between Al electrodes. The current-voltage (I-V) characteristics showed that the bipolar resistive switching is mainly attributed to bulk trap-dominated SCLC model. From the HRTEM and STEM point EDS analyses, it is suggested that the traps are deeply associated with positive space charges induced by the electrochemical migration of PSS⁻ chains into Al top electrode due to strong oxidation of Al metal.

**Acknowledgments**

This work was supported by the Next-generation Non-volatile Memory Program of the Ministry of Knowledge Economy, Korea.

**Figure Captions**

Fig. 1. Typical current density-voltage (J-V) characteristics of Al/PEDOT:PSS/Al device. The chemical structure of PEDOT:PSS is shown in the left bottom inset. The schematic diagram of the device is also illustrated in the right bottom inset.

Fig. 2. A double-logarithmic plot of J-V characteristics of Al/PEDOT:PSS/Al device in (a) negative bias and (b) positive bias regions. The bias sweep sequence is indicated by the arrows.

Fig. 3. Cross-sectional TEM images of Al/PEDOT:PSS/Al stacked structure. (a) BF image clearly showing each layers. (b) and (c) are enlarged HRTEM images obtained at both interfaces.

Fig. 4. (a) Cross-sectional ADF STEM image of Al/PEDOT:PSS/Al structure presenting atomic Z-contrast. (b), (c), and (d) represent point EDS spectra obtained from positions marked with red circles in the ADF image, respectively.

Fig. 5. Schematic diagrams showing the change of morphology and composition in PEDOT:PSS polymer film (a) before and (b) after top Al deposition.

Figure 1.

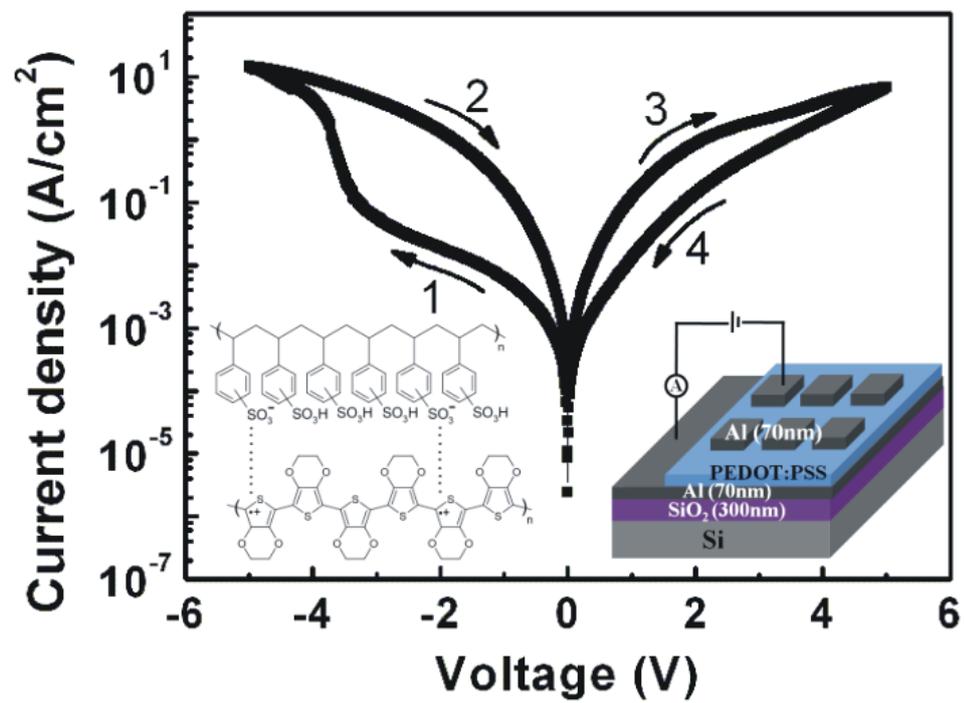

Figure 2.

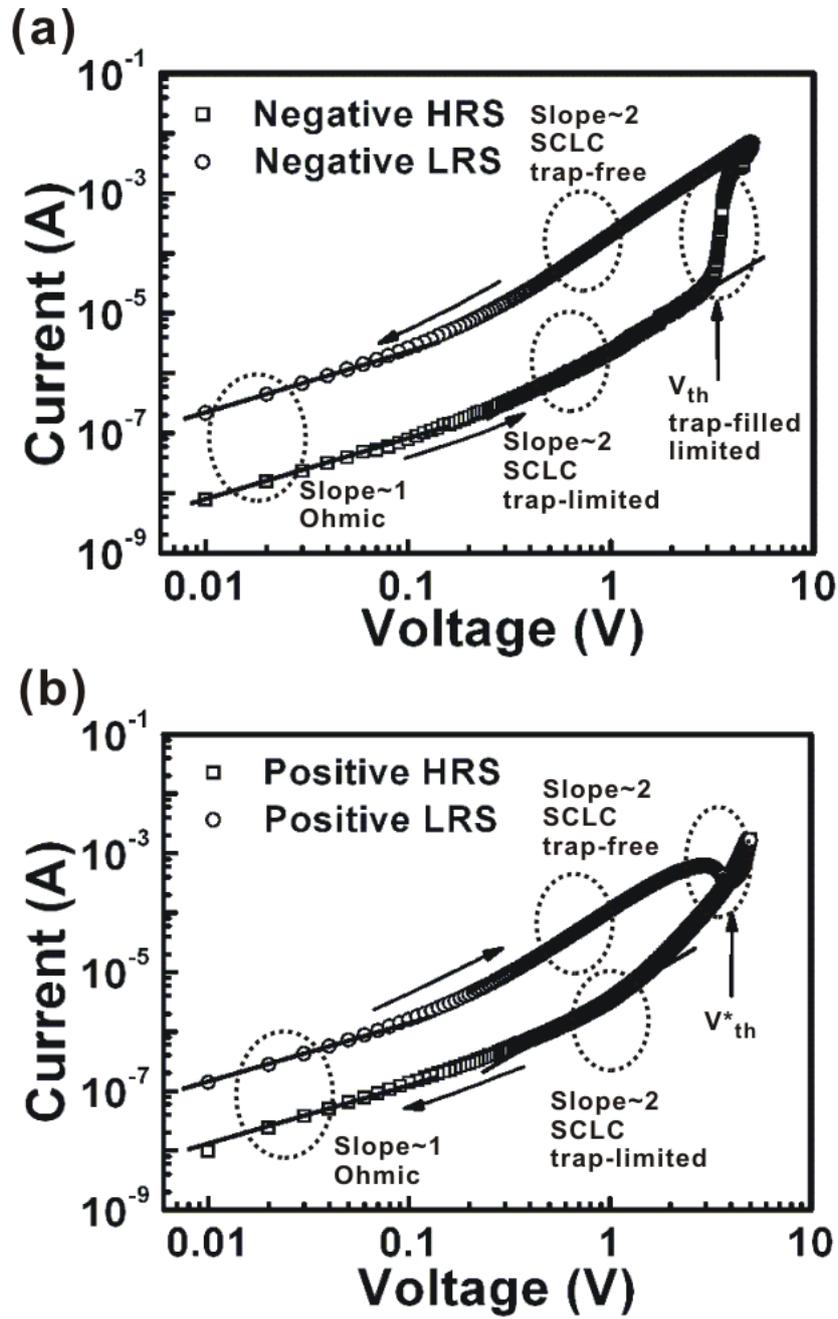

Figure 3.

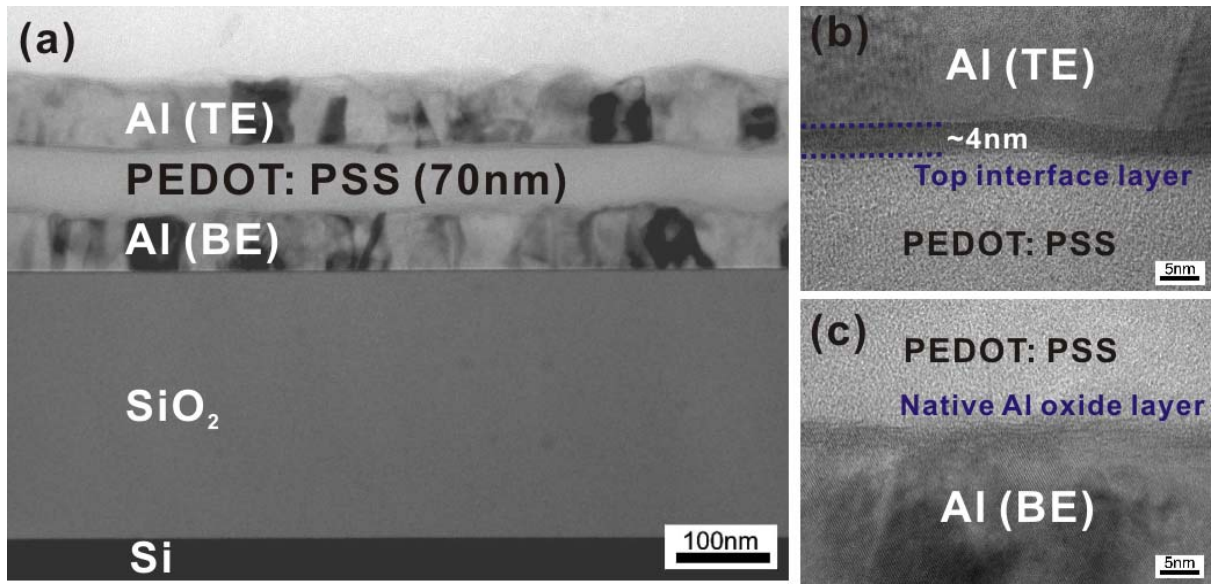

Figure 4.

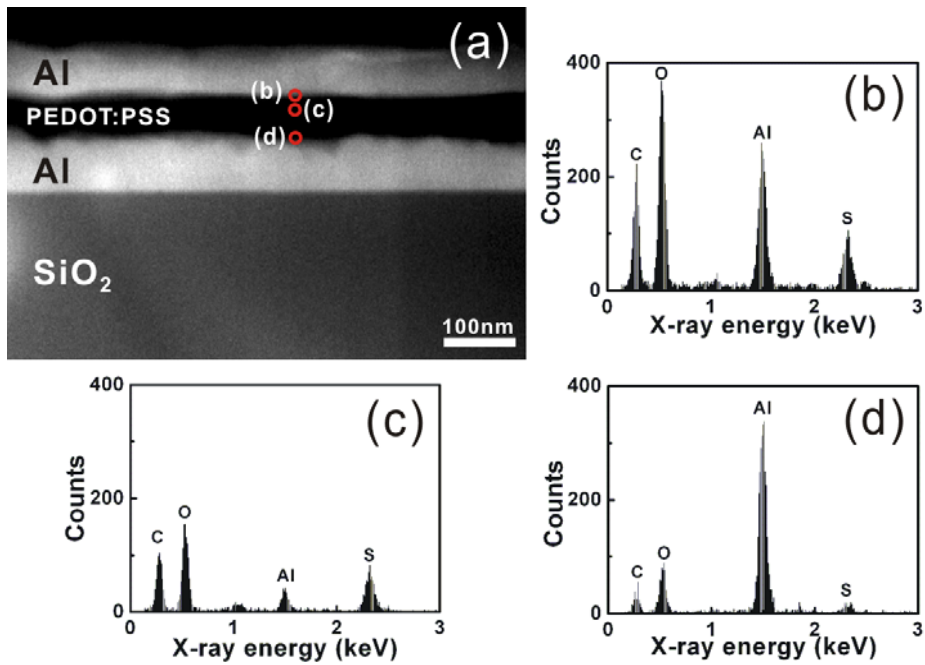

Figure 5.

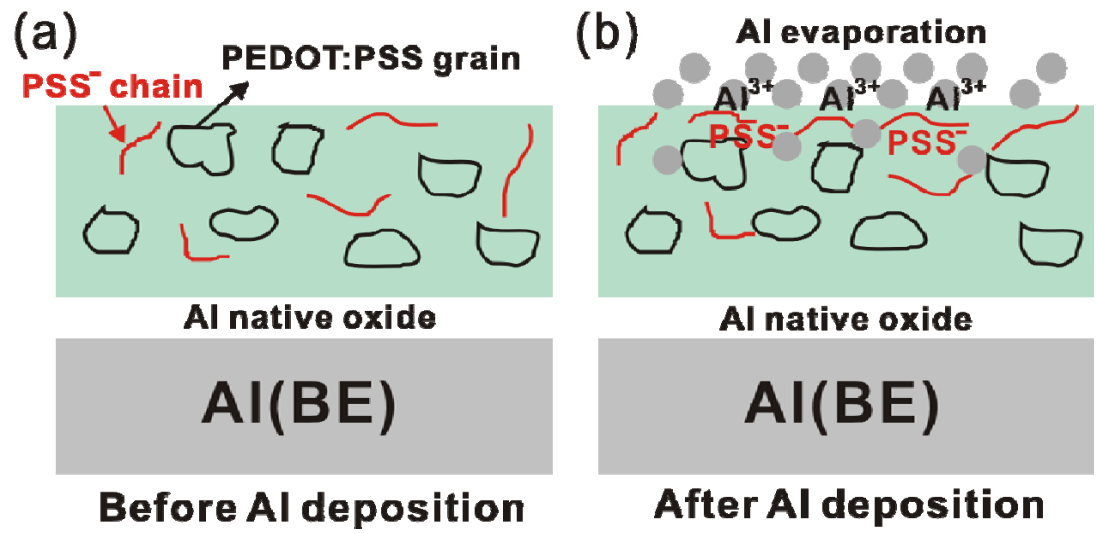